\documentclass[aps,prl,showpacs,twocolumn,superscriptaddress]{revtex4}
\usepackage{latexsym}
\usepackage{amssymb}
\usepackage{amsmath}
\usepackage{graphicx}
\begin{document}
\title{Inverse Spin Hall Effect by Spin Injection}
\author{S. Y. Liu}
\email{liusy@mail.sjtu.edu.cn}
\affiliation{Department of Physics,
Shanghai Jiaotong University, 1954 Huashan Road, Shanghai 200030,
China}
\author{ Norman J. M. Horing}
\affiliation{Department of Physics and Engineering Physics, Stevens
Institute of Technology, Hoboken, New Jersey 07030, USA}
\author{X. L. Lei}
\affiliation{Department of Physics, Shanghai Jiaotong University,
1954 Huashan Road, Shanghai 200030, China}
\begin{abstract}
Motivated by a recent experiment[Nature {\bf 442}, 176 (2006)], we
present a quantitative microscopic theory to investigate the inverse
spin-Hall effect with spin injection into aluminum considering both
intrinsic and extrinsic spin-orbit couplings using the
orthogonalized-plane-wave method. Our theoretical results are in
good agreement with the experimental data. It is also clear that the
magnitude of the anomalous Hall resistivity is mainly due to
contributions from extrinsic skew scattering, while its spatial
variation is determined by the intrinsic spin-orbit coupling.

\end{abstract}

\pacs{ 72.25.Ba, 72.25.Hg, 71.70.Ej}

\maketitle In the presence of an applied electric field, spin-orbit
(SO) coupling can cause deflection of electrons with opposite spins
in opposite directions, resulting in a nonvanishing transverse spin
current. This so-called spin-Hall effect (SHE), which was first
predicted by D'yakonov and Perel in 1971\cite{DP} and was recently
emphasized by Hirsch\cite{HS}, has attracted a great deal of
experimental and theoretical interest\cite{Review} because of its
potential applications in Spintronics. Depending upon its origin,
the SHE is classified into two types: intrinsic\cite{DP,HS,Halperin}
and extrinsic SHE\cite{Zhang,Sinova}. Experimental observations of
SHE have also been reported recently\cite{Kato,
Wunderlich,Zhao,Shen}.

The SO coupling mechanism can also lead to a reciprocal SHE
effect, the "inverse spin-Hall effect" (ISHE), which was predicted
by Hirsch\cite{HS}. When a purely longitudinal spin current is
applied, the electrons with opposite spins, flowing in opposing
longitudinal directions, move toward the same transverse side of
the sample due to SO interaction, resulting in charge
accumulation. Recently, Valenzuela and Tinkham presented a first
clear ISHE observation\cite{Nature}. By a nonlocal spin injection
into metallic aluminum through a ferromagnetic/nonmagnetic contact
(the configuration is illustrated in Fig.\,1), they observed a
finite Hall resistivity of order of milli-Ohms over a
spin-diffusion length of the order of $\mu$m.

It is well known that spin\cite{Spin-R} and charge\cite{Sorbello}
relaxations in aluminum are strongly influenced by band-structure
anomalies which arise mainly from several accidental degeneracy
points (near W points) in the Brillouin zone (BZ). Therefore, in an
analysis of the ISHE by spin injection in aluminum, the momentum
dependence of the scattering should be taken into account. Earlier
work by Zhang analyzing the SHE in the presence of spin diffusion
treated only momentum-independent relaxation\cite{ZhangS}. In this
Letter, we provides a realistic, quantitative theory for ISHE by
spin injection in aluminum considering the momentum dependencies of
the scattering rates and of the side-jump and skew scattering
contributions to the anomalous Hall current (AHC). Both the
intrinsic and extrinsic SO couplings are taken into account and the
electronic bands and states are determined by the
orthogonalized-plane-wave (OPW) method. The resulting Hall
resistivities are in good agreement with the experimental data. It
is also clear that, with spin injection into aluminum, the spatial
variation of AHC is determined by the intrinsic SO coupling while
its magnitude is mainly due to contributions from the extrinsic skew
scattering.

\begin{figure}
\includegraphics [width=0.45\textwidth,clip] {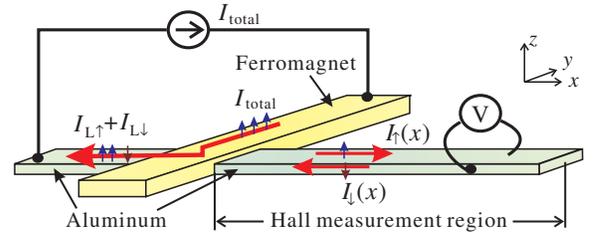}
\caption{(Color online) Measurement apparatus configuration for the
spin injection experiment of Ref.\,\cite{Nature}. A net current
$I_{\rm total}$ is injected through aluminum (Al) via a
ferromagnetic contact and it flows away from the Hall measurement
region. Near the contact, {\it i.e.} at $x=0$, $I_{\rm
total}=I_{L\uparrow}+I_{L\downarrow}$,
$I_{L\downarrow}=I_{\downarrow}(x=0)$, and
$I_{\uparrow}(x=0)+I_{\downarrow}(x=0)=0$, resulting in a vanishing
net electric current for $x>0$ and a finite spin current,
$I_{\uparrow}(x)-I_{\downarrow}(x)$, over a distance of order of the
spin-diffusion length.} \label{fig1}
\end{figure}

In general, the SO interaction induced by ionic fields, {\it i.e.}
the intrinsic SO coupling, may affect both the energy bands and the
states of electrons in solids. However, in aluminum, its effect on
the electronic band structure can be ignored. Hence, in our
treatment, the energy band structure of electrons is determined by
the standard OPW method in the absence of SO interaction, while the
electronic Bloch states are a mixture of spin up and spin down
species: $\psi^{(\mu)}_{n{\bf p}}({\bf r})=\sum_{\bf G}\left
[a^{(\mu)}_{n{\bf p}}({\bf G})|\mu>+b^{(\mu)}_{n{\bf p}}({\bf
G})|\bar\mu>\right ]\exp[i({\bf p}-{\bf G})\cdot {\bf r}]$, with
$\mu=1,2$ representing spinors ($\bar \mu=3-\mu$),  ${\bf p}$ as
lattice momentum confined to the first BZ, $n$ as band index, and
${\bf G}$ as reciprocal lattice vector. The coefficient
$a^{(\mu)}_{n{\bf p}}({\bf G})$ is determined by the pseudopotential
form of the Schr\"odinger equation, while $b^{( \mu)}_{n{\bf
p}}({\bf G})$ is obtained from first-order perturbation theory by
taking account of the SO part of the pseudopotential, $\lambda\hat
{\bf L}\cdot \hat {\bf S}{\cal P}_1$ ($\lambda$ is the SO coupling
constant and ${\cal P}_1$ is the operator projecting on the orbital
angular momentum state $l=1$).

In the spirit of the pseudopotential method, the electron-impurity
scattering potential, $V({\bf p},{\bf k})$, corresponding to
scattering of an electron from state ${\bf p}$ to state ${\bf k}$,
is the difference between the impurity- and host-atom form factors
with a structure factor correction for lattice distortion around the
substitutional impurities. It involves not only a spin-independent
part $V_0({\bf p},{\bf k})$ but also a SO part $V_{\rm SO}({\bf
p},{\bf k})$: $V({\bf p},{\bf k})=V_0({\bf p},{\bf k})+V_{\rm
SO}({\bf p},{\bf k})$. In general, the elements of $V_{\rm SO}({\bf
p},{\bf k})$, $V^{\mu\nu}_{\rm SO}({\bf p},{\bf k})$, take the form,
$V^{\mu\nu}_{\rm SO}({\bf p},{\bf k})\equiv A(p,k){\bf p}\times {\bf
k}\cdot <\nu|{\vec \sigma}|\mu>$ with $\vec \sigma$ as the Pauli
matrices, and $A(p,k)$ depending only on the magnitudes of the
electron momenta ${\bf p}$ and ${\bf k}$.

In the presence of spin diffusion, the transport properties of
solids may be described by a Wigner distribution function $\hat
\rho(n,{\bf p},{\bf R})$, which is a $2\times 2$ diagonal matrix
depending on the band index $n$, and on both Wigner coordinates:
microscopic electron momentum ${\bf p}$ and macroscopic space
coordinate ${\bf R}\equiv (x,y,z)$. Under steady state conditions,
$\hat \rho(n,{\bf p},{\bf R})$ obeys a kinetic (Boltzmann) equation
given by
\begin{eqnarray}
\frac{\bf p}{m}\cdot{\bf \nabla}_{\bf R}\hat \rho_{\mu\mu}(n,{\bf
p},{\bf R})=\sum_{{\bf p},{\bf k},n'}\left \{W_{\mu\mu}^{nn'}({\bf
p},{\bf k}) \right .\nonumber\\
\times [\hat \rho_{\mu\mu}(n',{\bf k},{\bf R})-\hat
\rho_{\mu\mu}(n,{\bf p},{\bf R})] \nonumber\\
\left . + W_{\mu\bar\mu}^{nn'}({\bf p},{\bf k})[\hat
\rho_{\bar\mu\bar\mu}(n',{\bf k},{\bf R})-\hat \rho_{\mu\mu}(n,{\bf
p},{\bf R})]\right \}.\label{KE}
\end{eqnarray}
Here, $W_{\mu\mu}^{nn'}({\bf p},{\bf k})$ and
$W_{\mu\bar\mu}^{nn'}({\bf p},{\bf k})$, respectively, are the
non-spin-flip and spin-flip scattering rates. They can be obtained
by analyzing the matrix elements of the scattering potential:
\begin{eqnarray}
W_{\mu\mu}^{nn'}({\bf p},{\bf k})=2\pi \delta(\varepsilon_{n {\bf
p}}-\varepsilon_{n' {\bf k}})\,\,\,\,\,\,\,\,\,\,\,\,\,\,\,\,\,\,\,\,\,\,\,\,\,\,\,\,\,\,\,\,\,\,\,\,\,\,\,\,\,\,\,\,\,\,\,\,\,\,\,\nonumber\\
\times\left |\sum_{{\bf G},{\bf G}'}\left \{\left [a^{(\mu)}_{n'{\bf
k}}({\bf G}')+b^{(\bar\mu)}_{n'{\bf k}}({\bf G}')\right ]^*\right
.\right.\,\,\,\,\,\,\,\,\,\,\,\,\,\,\,\,\,\,\,\,\,\,\,\,\,\,\,\,\,\,\,\,\,\,\,\,\nonumber\\
\times\left [a^{(\mu)}_{n{\bf p}}({\bf G})+b^{(\bar\mu)}_{n{\bf
p}}({\bf G})\right ]V_0({\bf p}-{\bf
G},{\bf k}-{\bf G'})\nonumber\\
\left. \left .+[a^{(\mu)}_{n'{\bf k}}({\bf G}')]^*a^{(\mu)}_{n{\bf
p}}({\bf G})V_{\rm SO}^{\mu\mu}({\bf p}-{\bf G},{\bf k}-{\bf
G'})\right \}\right |^2
\end{eqnarray}
and
\begin{eqnarray}
W_{\mu\bar\mu}^{nn'}({\bf p},{\bf k})=2\pi \delta(\varepsilon_{n
{\bf p}}-\varepsilon_{n' {\bf k}})\,\,\,\,\,\,\,\,\,\,\,\,\,\,\,\,\,
\,\,\,\,\,\,\,\,\,\,\,\,\,\,\,\,\,\,\,\,\,\,\,\,\,\,\,\,\,\,\,\,\,\,\nonumber\\
\times\left |\sum_{{\bf G},{\bf G}'}\left \{\left (a^{(\mu)}_{n{\bf
p}}({\bf G})[b^{(\bar \mu)}_{n' {\bf k}}({\bf
G}')]^*\right .\right .\right .\,\,\,\,\,\,\,\,\,\,\,\,\,\,\,\,\,
\,\,\,\,\,\,\,\,\,\,\,\,\,\,\,\,\,\,\,\,\,\,\,\nonumber\\
\left . +b^{(\mu)}_{n{\bf p}}({\bf G})[a^{(\bar \mu)}_{n' {\bf
k}}({\bf G}')]^*\right )V_0({\bf p}-{\bf G},{\bf k}-{\bf G'})\nonumber\\
\left .\left . +a^{(\mu)}_{n{\bf p}}({\bf G})[a^{(\bar \mu)}_{n'{\bf
k}}({\bf G}')]^*V_{\rm SO}^{\mu\bar\mu}({\bf p}-{\bf G},{\bf k}-{\bf
G'})\right \}\right |^2,
\end{eqnarray}
with $\varepsilon_{n{\bf p}}$ as the energy dispersion relation for
$n$-th band electrons. Note that in Eqs. (2) and (3) both the
intrinsic and extrinsic SO couplings are involved. The last term on
the right-hand side of Eq.\,(\ref{KE}) leads to an exchange of
electrons with different spins, causing spin relaxation. This
so-called Elliot-Yafet spin-relaxation mechanism\cite{E,Y} in
aluminum has been carefully investigated by Fabian and Das Sarma
taking account of electron-phonon scattering\cite{Spin-R}. They
found that spin relaxation in aluminum is determined mainly by
several accidental degeneracy points near the W points in BZ.
%\end{widetext}

It has already been demonstrated that intrinsic SO coupling can make
a nonvanishing contribution to AHC if the off-diagonal elements of
the distribution function are finite\cite{Liu2}. However, in the
spin injection experiment, the distribution function is essentially
diagonal since the off-diagonal driving forces, such as the electric
field and Zeeman energy splitting, are negligible. In particular, in
the experiment of Ref.\cite{Nature}, there is no external driving
electric field in the Hall measurement region, while the Zeeman
energy splitting for the maximum applied magnetic field, B = 3.5T,
is only 0.2meV, much smaller than the free-electron Fermi energy of
aluminum, $E_{\rm F}^0 = 11.7$\,eV. Hence, to determine AHC, one
only needs to consider the extrinsic SO coupling.

The extrinsic SO interaction makes nonvanishing contributions to AHC
through two mechanisms: a side-jump process proposed by
Berger\cite{Berger} and also skew scattering given by
Smit\cite{Smit}. The side-jump AHC arises from a sidewise shift of
the center of the electron wave packet, while the skew scattering
contribution corresponds to an anisotropic enhancement of the wave
packet due to electron-impurity scattering  described in the second
Born approximation\cite{Hugon}. Obviously, the analysis of spin
injection in aluminum calls for the formulation of these two
mechanisms of AHC in terms of a distribution function. In previous
studies, these two AHC mechanisms were described in the framework of
a momentum-independent relaxation time\cite{Crepieux,Review2}, or in
linear response theory\cite{Liu}, and hence the existing results can
not be applied in the present study of the ISHE by spin injection in
aluminum.

To analyze the side-jump process, we begin with the anomalous term
of the current operator, which is associated with function
$<\psi^+_{\mu} ({\bf r}_1,t)\psi_{\mu} ({\bf r},t)>$ ($\psi_{\mu}
({\bf r},t)$ and $\psi^+_{\mu} ({\bf r},t)$ are electron field
annihilation and creation operators, respectively). Expanding the
statistical average involved in this function to first order of the
electron-impurity interaction and applying the Langreth
algebra\cite{Jauho} taken jointly with the generalized Kadanoff-Baym
ansatz\cite{GKBA,GKBA1}, we finally arrive at the form of the
steady-state current, ${\bf J}^{\rm SJ}({\bf R})$,  in the
lowest-order gradient expansion of the Wigner coordinate ${\bf R}$:
\begin{eqnarray}
J^{\rm SJ}_\alpha({\bf R})=-i\pi e N_i\sum_{\substack{{\bf p},{\bf q}\\
\mu,n,n'}}\sum_{\substack{{\bf G}_1,{\bf G}_2\\{\bf G}_3,{\bf
G}_4}}(-1)^\mu \bar A^{(\mu)}(n{\bf p},n'{\bf k},{\bf G}_1,{\bf
G}_2)\nonumber\\
\times \bar V_0^{(\mu)}(n'{\bf k},n{\bf p},{\bf G}_3,{\bf G}_4)
 \varepsilon_{\alpha\beta z}[({\bf k}-{\bf G}_2)-({\bf p}-{\bf
G}_1)]_\beta\nonumber\\
\times \delta(\varepsilon_{n'{\bf k}}-\varepsilon_{n{\bf p}})
 [\hat\rho_{\mu\mu}(n,{\bf p},{\bf R})-\hat\rho_{\mu\mu}(n',{\bf
 k},{\bf R})],\,\,\,\,\,\,\,\,\,\,\,\,\,\,\,\,\,\,\label{SJ}
\end{eqnarray}

with $\bar A^{(\mu)}(n{\bf p},n'{\bf k},{\bf G}_1,{\bf G}_2)\equiv
[a^{(\mu)}_{n'{\bf k}}({\bf G}_2)]^*a^{(\mu)}_{n{\bf p}}({\bf
G}_1)A(|{\bf p}-{\bf G}_1|,|{\bf k}-{\bf G}_2|)$, $\bar
V_0^{(\mu)}(n{\bf p},n'{\bf k},{\bf G}_1,{\bf G}_2)\equiv
[a^{(\mu)}_{n'{\bf k}}({\bf G}_2)]^*a^{(\mu)}_{n{\bf p}}({\bf
G}_1)V_0({\bf p}-{\bf G}_1,{\bf k}-{\bf G}_2)$, and
$\varepsilon_{\alpha\beta z}$ as the totally antisymmetric tensor.

The skew scattering AHC, ${\bf J}^{\rm SS}({\bf R})$, is associated
with the additional SO term of the distribution function in the
second Born approximation. Substituting this term into the non-SO
part of current operator, in the first order of SO coupling constant
$\lambda$, ${\bf J}^{\rm SS}({\bf R})$ takes the form,
%\begin{widetext}
\begin{eqnarray}
{\bf J}^{\rm SS}({\bf R})=i 4\pi^2 eN_i\sum_{\substack{{\bf p},{\bf
k},{\bf q}\\ \mu, n,n',n''}}\sum_{\substack{{\bf G}_1,{\bf G}_2,{\bf
G}_3\\ {\bf G}_4,{\bf G}_5,{\bf G}_6}}{\bf v}_{n\bf
p}\,\,\,\,\,\,\,\,\,\,\,\,\,\,\,\,\,\,\,\,\,\,\,\,\,\,\,\,\nonumber\\
\times(-1)^\mu\delta(\varepsilon_{n'{\bf k}}-\varepsilon_{n{\bf
p}})\delta(\varepsilon_{n''{\bf q}}-\varepsilon_{n{\bf
p}})\,\,\,\,\,\,\,\,\,\,\,\,\,\,\,\,\,\,\,\,\,\,\,\,\,\,\,\,\,\,\,\,\,\,\,\,\nonumber\\
\times \left \{\bar A^{(\mu)}(n{\bf p},n'{\bf k},{\bf G}_1,{\bf
G}_2)V_0^{(\mu)}(n'{\bf k},n''{\bf q},{\bf G}_3,{\bf G}_4)
\right .\nonumber\\
\times V_0^{(\mu)}(n''{\bf q},n{\bf p},{\bf G}_5,{\bf
G}_6)\varepsilon_{\alpha\beta z}[{\bf p}-{\bf
G}_1]_\alpha[{\bf k}-{\bf G}_2]_\beta\nonumber\\
+\left .\left [
\begin{array}{ccc}
n{\bf p}\rightarrow n'{\bf k}\\
n'{\bf k}\rightarrow n''{\bf q}\\
n''{\bf q}\rightarrow n{\bf p}
\end{array}
\right ]+\left [
\begin{array}{ccc}
n{\bf p}\rightarrow n''{\bf q}\\
n'{\bf k}\rightarrow n{\bf p}\\
n''{\bf q}\rightarrow n'{\bf k}
\end{array}
\right ]\right \} \,\,\,\,\,\,\,\,\,\,\,\,\,\,\,\,\,\,\,\nonumber\\
\times \hat \rho_{\mu\mu}(n',{\bf k},{\bf R})\tau(n',{\bf
k}),\,\,\,\,\,\,\,\,\,\,\,\,\,\,\,\,\,\,\,
\,\,\,\,\,\,\,\,\,\,\,\,\,\,\,\,\,\,\,
\,\,\,\,\,\,\,\,\,\,\,\,\,\,\,\,\,\,\,\label{SS}
\end{eqnarray}
with ${\bf v}_{n\bf p}\equiv {\rm d}\varepsilon_{n{\bf p}}/{\rm
d}{\bf p}$ as the electron velocity and $N_i$ as the impurity
density. $\tau(n',{\bf k})$ is the transport relaxation time, which
has been carefully investigated in Ref.\,\cite{Sorbello}. The last
two terms in curly brackets of Eq.\,(\ref{SS}) are obtained from the
first term by following the replacement rules indicated in the
square brackets.

It should be noted that the side-jump and skew scattering
contributions to AHC, $J_{\alpha}^{\rm SJ}({\bf R})$ and
$J_{\alpha}^{\rm SS}({\bf R})$, exist in conjunction with a
nonvanishing longitudinal spin current, rather than the net electric
current. In a system with a longitudinal charge current but without
magnetization, the electrons with opposite spins distribute equally
and the factor $(-1)^{\mu}$ in Eqs. (\ref{SJ}) and (\ref{SS}) leads
to vanishing of the AHC. However, when a purely longitudinal spin
current is present, we have $\hat \rho^{(1)}_{\mu\mu}(n',{\bf
k},{\bf R})=-\hat\rho^{(1)}_{\bar\mu\bar\mu}(n',{\bf k},{\bf R})$
with $\hat \rho^{(1)}(n',{\bf k},{\bf R})$ as the deviation from the
equilibrium Fermi distribution. Hence, the signs of contributions to
Hall current from electrons having opposite spins are the same,
resulting in a finite transverse charge current.

We perform a numerical calculation to investigate the ISHE in the
spin injection experiment illustrated in Fig.\,1. The electronic
band structure of aluminum is evaluated by a {\it modified} 4 OPW
method, in which the 15 shortest reciprocal lattice vectors are
considered but the pseudopotentials, except for Ashcroft's $V_{111}$
and $V_{200}$\cite{Ashcroft}, are assumed to vanish. This treatment
enables us to obtain a continuous electron velocity when the
electron momentum crosses from one symmetry section of the fcc BZ
into another. The sums over all electron momenta in the first BZ are
performed by the tetrahedron method\cite{MacDonald}. Further, we
assume that the electron-impurity interaction arises from defects,
which may be introduced in the process of material fabrication. In
this way, the scattering potential is just given by the form factor
of aluminum but with the opposite sign. In the calculations, the
spin-independent part of scattering potential is described by a
screened Ashcroft form factor\cite{Wiser}, while the SO part is
chosen from Ref.\cite{Anderson} with SO coupling constant
$\lambda=2.4\times 10^{-3}$\,a.u.\cite{Spin-R}. To determine the
transverse Hall current, $J_y\equiv J_y^{\rm SJ}+J_y^{\rm SS}$, and
the longitudinal spin current, $J^z_x(x)\equiv J_{\uparrow
x}(x)-J_{\downarrow x}(x)$ with $J_{\mu x}(x)=\sum_{n,\bf p}e({\bf
v}_{n{\bf p}})_x\hat\rho_{\mu\mu}({\bf p},x)$,  we solve the
Boltzmann equation, Eq.\,(\ref{KE}), using the
Fermi-surface-harmonic-expansion method\cite{SHE}, assuming an
initial deviation from the equilibrium Fermi distribution,
$\hat\rho^{(1)}_{\mu\mu}(n,{\bf p},x=0)$, of form:
$\hat\rho^{(1)}_{\mu\mu}(n,{\bf p},x=0)=(-1)^\mu e {\bf E}_{\rm
c}\cdot {\bf v}_{n{\bf p}}\tau_{n{\bf p}}n'_{\rm
F}(\varepsilon_{n{\bf p}})$ with $n'_{\rm F}(\varepsilon)$ as the
derivative of the Fermi function. The effective electric field ${\bf
E}_c$ is obtained by considering the initial spin-up current
densities injected into the right-hand side of the aluminum at $x=0$
in the experiment of Ref.\,\cite{Nature}:
$J_{{\uparrow}x}(x=0)=3.75$ or $1.8\times 10^9$\,A/m$^{2}$ for
samples with thicknesses $t=12$ or $25$\,nm (in the experiment,
$I_{\rm total}=50$\,$\mu$A, the sample widths are $w=400$\,nm and
the spin-polarization,
$P\equiv(I_{L\uparrow}-I_{L\downarrow})/I_{\rm total}$ at $x=0$, is
$0.28$). The impurity densities are determined from the mobilities
of the experimental samples.

\begin{figure}
\includegraphics [width=0.45\textwidth,clip] {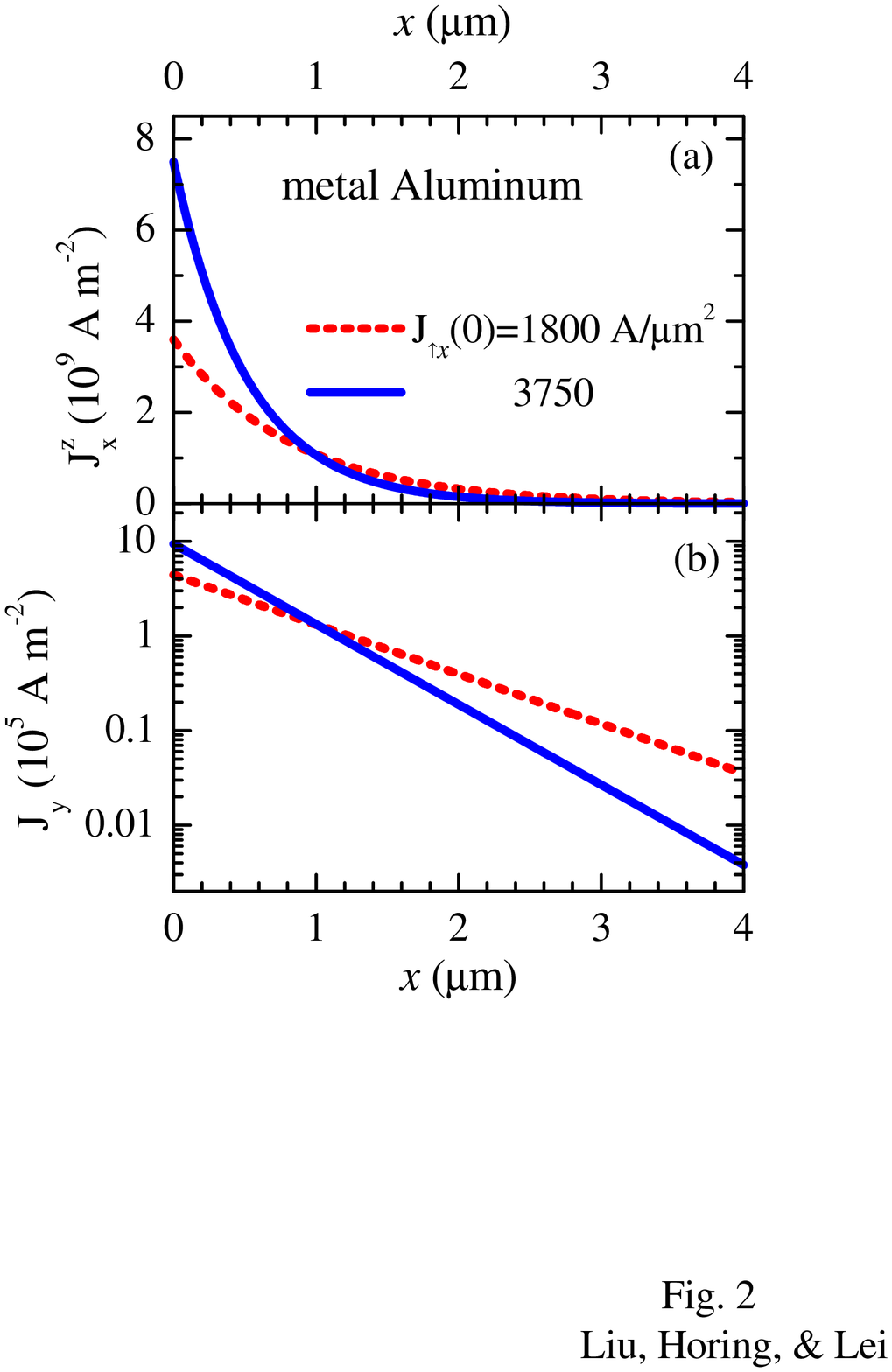}
\caption{(Color online) Coordinate dependencies of the longitudinal
spin current, $J_x^z$ (a), and anomalous Hall current, $J_y$ (b), in
the spin injection experiment for different initial values of
$J_{\uparrow x}(x)$: $J_{\uparrow x}(0)=3750$, and
$1800$\,A/$\mu$m$^2$.} \label{fig2}
\end{figure}

In Fig.\,2, we plot the longitudinal spin current $J_x^{z}$ and the
AHC $J_y$ as functions of coordinate $x$ for the two different
values of $J_{\uparrow x}(x=0)$. It is evident that with increasing
$x$, $J_x^z(x)$ and $J_y(x)$ decrease exponentially: $J_x^z(x)$,
$J_y(x)$ $\sim \exp(-x/x_{sd})$ with $x_{sd}$ as the spin-diffusion
length. For aluminum, we find $x^{\rm theory}_{sd}=830$\,nm for
$t=25$\,nm, and $x^{\rm theory}_{sd}=512$\,nm for $t=12$\,nm, which
are in good agreement with the experimental results: the
experimental $x_{sd}$ values are $x^{\rm exp}_{sd}=735$\,nm for
$t=25$\,nm and $x^{\rm exp}_{sd}=490$\,nm for $t=12$\,nm.

From Fig.\,2 we see that the AHC exists but is relatively small: it
is almost five orders of magnitude smaller than $J_x^z(x)$.
Nevertheless, the corresponding Hall voltage can be detected. We
calculated the Hall resistivities for the two initial values of
$J_{\uparrow x}(x)$ and compared them with the experimental data.
The results are plotted in Fig.\,3. Our theoretical results, without
any adjustable parameters, are in good agreement with the data. For
$t=12$\,nm (or $J_{\uparrow x }(0)=3750$\,A/m$^{2}$), our
theoretical values for $R_{\rm AHE}$  agree perfectly with the data.
For $t=25$\,nm (or $J_{\uparrow x }(0)=1800$\,A/m$^{2}$), the
calculated results are a bit larger than those obtained
experimentally. This may be due to the anisotropy of the
experimental sample, which could lead to a different $\lambda$-value
than the one we use here. Note that the sign of the calculated Hall
voltage also agrees with that found experimentally: the positive
spin current, $J_{x }^z(x)$, polarized along the positive
$z$-direction produces a negative Hall electric field.

\begin{figure}
\includegraphics [width=0.45\textwidth,clip] {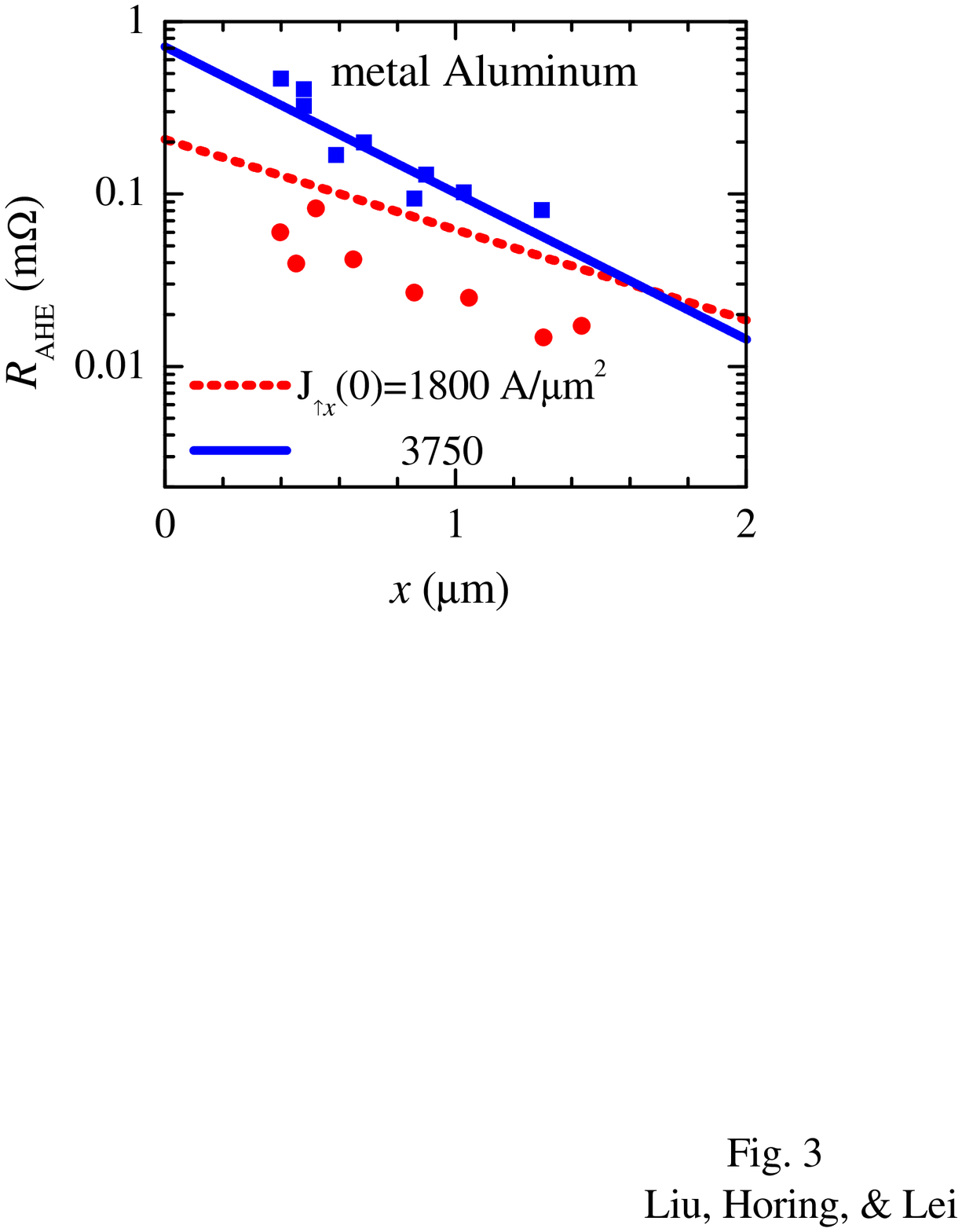}
\caption{(Color online) Anomalous Hall resistivities, $R_{\rm AHE}$,
as functions of coordinate $x$ for different initial values of
$J_{\uparrow x}(x)$: $J_{\uparrow
x}(0)=3750$,\,$1800$\,A/$\mu$m$^2$. The solid and dotted lines are
our theoretical results, and the squares (circles) are the
experimental data for Al thicknesses $t=12$ (25) nm
(Ref.\,\cite{Nature}).} \label{fig3}
\end{figure}

Our numerical calculation also shows that the contributions to AHC
are dominated by skew scattering. The side-jump AHC contribution is
only about 3.5\% of the skew scattering one. This is due to the fact
that the mobilities of the samples involved are small and the
impurities are relatively dense: $N_i$ is about 1\% of the electron
density. Note that the signs of both the side-jump and skew
scattering AHC contributions are the same.

We also find that the effect of the extrinsic SO coupling on the
diffusion length vanishes. This holds valid for materials with space
inversion symmetry (SIS). In the case of weak SO coupling, $a_{n{\bf
p}}^{(\mu)}$ is real: $a_{n{\bf p}}^{(\mu)}=[a_{n{\bf
p}}^{(\mu)}]^*$, and SIS leads to $a_{n{\bf p}}^{(\bar\mu)}\equiv
a_{n,-{\bf p}}^{(\mu)}=a_{n{\bf p}}^{(\mu)}$. Hence, the extrinsic
parts of the scattering rates are asymmetric under the momentum
inversion operation, resulting in a vanishing contribution to the
diffusion length. This implies that the spatial dependence of AHC is
determined only by the intrinsic SO coupling since $J_{\alpha}^{\rm
SJ}({\bf R})$ and $J_{\alpha}^{\rm SS}({\bf R})$ depend on ${\bf R}$
only through the distribution function.

In conclusion, we have presented a fully microscopic theory to
analyze the ISHE observed in the spin-injection experiment in
aluminum taking account of both the intrinsic and extrinsic SO
couplings. Employing the OPW method, the momentum-dependent
scattering rates of the Boltzmann equation were treated carefully
and the side-jump and skew scattering contributions to the AHC were
formulated in terms of the distribution function. Performing
realistic calculations for aluminum, we have determined anomalous
Hall resistivities that are in good agreement with the experimental
data.

The authors would like to thank Dr. S. O. Valenzuela for providing
the details of their experiment and its setup. This work was
supported by the Youth Scientific Research Startup Funds of SJTU, by
projects of the National Science Foundation of China and the
Shanghai Municipal Commission of Science and Technology, and by the
Department of Defense through the DURINT program administered by the
US Army Research Office, DAAD Grant No. 19-01-1-0592.

\end{document}